\documentclass[prd,aps,superscriptaddress,10pt,nobibnotes,notitlepage,twocolumn,longbibliography,nofootinbib,preprintnumbers]{revtex4-2}

\usepackage[normalem]{ulem}
\usepackage{sidecap}
\usepackage[latin1]{inputenc}
\usepackage{graphicx}
\usepackage{float}
\usepackage{latexsym}
\usepackage{graphicx}
\usepackage{amssymb}
\usepackage{amsmath,mathtools}
\usepackage{amsfonts}
\usepackage{tikz} 
\usepackage{ragged2e}
\usepackage{stackengine}
\usepackage{scalerel}

\newcommand{\lowersub}{\smash[b]{\vphantom{\bigg|}}}

\usepackage[colorlinks=true,citecolor=blue,hyperfootnotes=false]{hyperref}

\usepackage{dcolumn}
\usepackage{textcomp}
\usepackage{xfrac}
\usepackage{slashed}
\usepackage{multirow}

\def\be{\begin{equation}}
\def\ee{\end{equation}}
\def\figs/B{B}
\def\bea{\begin{eqnarray}}
\def\eea{\end{eqnarray}}
\def\bg{\begin{eqnarray}}
\def\nd{\end{eqnarray}}

\def\ln{{\rm log}}

\def\beq{\begin{equation}}
\def\eeq{\end{equation}}

\usepackage{graphicx}
\usepackage{dcolumn}
\usepackage{bm}
\usepackage{hyperref}
\usetikzlibrary{decorations.pathmorphing}
\tikzset{snake it/.style={decorate, decoration=snake}}
\newcommand{\sint}[3]{\hspace{- #3 em} \underset{#1}{\overset{#2}{\int}}\hspace{- #3 em}}

\begin{document}

\preprint{MIT-CTP/5638}

\title{Finite-Temperature Instantons from First Principles}

\author{Thomas Steingasser}
 \email{tstngssr@mit.edu}
\affiliation{%
Department of Physics, Massachusetts Institute of Technology, Cambridge, MA 02139, USA }
\affiliation{Black Hole Initiative at Harvard University, 20 Garden Street, Cambridge, MA 02138, USA
}%

\author{Morgane K\"onig}
 \email{mkonig@mit.edu}
\affiliation{%
Department of Physics, Massachusetts Institute of Technology, Cambridge, MA 02139, USA
}%

\author{David I.~Kaiser}
 \email{dikaiser@mit.edu}
\affiliation{%
Department of Physics, Massachusetts Institute of Technology, Cambridge, MA 02139, USA
}%

\date{\today}

\begin{abstract}
We derive the finite-temperature quantum-tunneling rate from first principles. The tunneling rate depends on both temperature and time. We demonstrate that the relevant instantons should therefore be defined on a Keldysh-Schwinger contour, and discuss how the familiar Euclidean-time result arises from the limit of large physical times. We identify distinct behavior in the high- and low-temperature limits, incorporating effects from background fields. We construct a consistent perturbative scheme that incorporates large finite-temperature effects.
\end{abstract}

\maketitle

{\it \bf Introduction.} 
Tunneling is one of the most important examples of quantum phenomena~\cite{Kobzarev:1974cp,Coleman:1977py,Callan:1977pt,Devoto:2022qen,Linde:1980tt,Linde:1981zj}. Despite recent advances in our understanding of this process~\cite{Andreassen:2016cff,Andreassen:2016cvx,Andreassen:2017rzq,Khoury:2021zao,Steingasser:2022yqx,Chauhan:2023pur,Espinosa:2018hue,Espinosa:2018voj,Espinosa:2018szu,Espinosa:2019hbm,Espinosa:2022ofv,Witten:2010cx,Tanizaki:2014xba,Cherman:2014sba,Dunne:2015eaa,Bramberger:2016yog,Michel:2019nwa,Mou:2019gyl,Hertzberg:2019wgx,Ai:2019fri,Hayashi:2021kro,Nishimura:2023dky}, many of its aspects remain elusive, in particular for tunneling out of non-vacuum states, e.g., for finite temperatures.

Based on the zero-temperature limit for quantum systems and the rigorously derived transition-rate formula for classical systems at finite temperature~\cite{Langer:1967ax,Langer:1969bc,Bochkarev:1992rh,Boyanovsky:1993ki,Garny:2012cg,Ekstedt:2022tqk,Hirvonen:2021zej,Lofgren:2021ogg}, the tunneling rate $\Gamma$ per spatial volume $V$ of a false vacuum at finite temperature has long been conjectured to be given by the imaginary part of the system's free energy $F$~\cite{Linde:1980tt,Linde:1981zj} as
\begin{equation}\label{FE}
    \frac{\Gamma}{V} \simeq - 2 \ \text{Im} (F) = - 2 \, \text{Im} \left( -T \, \ln \sint{\varphi (0)=\varphi (\beta)}{}{1.2} \mathcal{D}\varphi \ 
     e^{-S_E [\varphi]} \right),
\end{equation}
in terms of the Euclidean action $S_E [\varphi]$. Throughout this article, $\beta \equiv T^{-1}$ denotes the inverse temperature. Eq.~\eqref{FE} amounts to imposing periodic boundary conditions along the Euclidean-time axis on the zero-temperature result of Ref.~\cite{Callan:1977pt}. 

The zero-temperature formula motivating this expression is plagued by several conceptual issues, which require non-trivial modifications if the well-established, leading-order result is to be recovered~\cite{Andreassen:2016cff,Andreassen:2016cvx,Andreassen:2017rzq}. Most important, the imaginary part of the bounce contribution to the path integral in Eq.~\eqref{FE} is canceled by contributions from other saddle points, implying a vanishing tunneling rate. Whereas these can be dealt with by artificially restricting the domain of the path integral through the so-called \textit{potential deformation method}~\cite{Andreassen:2016cvx}, the only motivation to do so is to recover Coleman's famous leading-order result. These cancellations do not depend on the size of the Euclidean time interval, so that the finite-temperature result inherits these issues. 

Another shortcoming of this approach is the role of time dependence. It is well-understood that the tunneling rate is \textit{a priori} a real-time dependent observable~\cite{Andreassen:2016cff,Andreassen:2016cvx}. In the zero-temperature case, the Euclidean-time dependence arises through a Wick rotation, and the apparent time independence through the limit of large times relative to the natural time-scale of the system of interest. How this procedure is consistent with the finite Euclidean-time interval in Eq.~(\ref{FE}) representing the system's temperature remains unclear.

{\bf Defining the Tunneling Rate.} To address these questions, we derive the tunneling rate $\Gamma$ from first principles following the \textit{direct approach} of Refs.~\cite{Andreassen:2016cff,Andreassen:2016cvx}. $\Gamma$ is defined through the change of probability to find the system within its classically accessible region $\mathcal F$, $\mathcal{P}_{\mathcal F}(t) = \mathcal{P}_{\mathcal F}(0) \cdot e^{- \Gamma t}$. This defines $\Gamma$ as
\begin{equation}
    \Gamma = - \frac{1}{\mathcal{P}_{\mathcal F}(t)} \frac{\text{d}}{\text{d} t}\mathcal{P}_{\mathcal F}(t)=\frac{1}{\mathcal{P}_{\mathcal F}(t)} \frac{\text{d}}{\text{d} t}\mathcal{P}_{\mathcal R}(t), \label{Gammadef}
\end{equation}
where ${\mathcal R}$ is the exterior of $\mathcal F$ in field space. See Fig.~\ref{fig:FRBarrier}.

Following Ref.~\cite{Shkerin:2021zbf}, we describe the system through its density matrix $\rho$. The probability to find the field in some region $\Omega$ of field space is then given by
\begin{equation}\label{fund}
    \mathcal{P}_\Omega (t)= \text{Tr}_\Omega\left[ \rho \right] (t) = \mathcal{N} \int_{\Omega} D \varphi \,\langle \varphi ,t | \rho | \varphi ,t \rangle .
\end{equation}
%
The right-hand side of Eq.~(\ref{Gammadef}) can be rewritten by inserting two partitions of unity:
\begin{equation}\label{PR1}
    \mathcal{P}_{\mathcal R}  \hspace{-0.2em} =  \sint{\mathcal{F}}{}{0.2} D\varphi_{i} D\varphi_{j} \langle\varphi_{i} |\rho|\varphi_j \rangle \sint{\mathcal R}{}{0.2} D\varphi_{f} \langle\varphi_j |\varphi_{f},t\rangle\langle\varphi_{f},t|\varphi_{i}\rangle .
\end{equation}
The first factor, $\rho_{ij}\equiv  \langle\varphi_{i} |\rho|\varphi_j \rangle$, encodes the initial state of the system. The second factor, $P_{ji} \equiv \int_{\mathcal R} D\varphi_{f} (...) $, captures the time evolution of the system. All states without an explicit time label are defined at $t=0$.

\begin{figure}
\begin{tikzpicture}[scale=6]
  \draw[->] (0,0) -- (1.15,0) node[right] {$\phi$};
  \draw[->] (0,0) -- (0,0.4) node[above] {$U[\phi]$};
\draw[domain=0:1.079,smooth,variable=\x,very thick] plot ({\x},{0.8*(\x)^2 - 0.6*(\x)^6});  
  \draw[gray!60,dashed] (0.81,0)--(0.81,0.36);
  \draw[gray!60,red,snake it, thick] (0,0.08)--(0.32,0.08);
  \draw[gray!60,red,snake it, thick] (0,0.2)--(0.51,0.2);
  \draw[gray!60,red,snake it, thick] (0,0.32)--(0.7,0.32);
  \draw (-0.01,-0.05) node [black] {$\phi_{\rm FV}$};
  \draw (0.8,-0.05) node [black] {$\phi_{\rm top}$};
  \draw (0.33,0.042) node [gray]  {$t=0$};
  \draw (1.02,0.042) node [gray]  {$t$};
  \draw (0.27,0.14) node [black]  {$\varphi_{i}$};
  \draw (1.1,0.14) node [black]  {$\Sigma_{\varphi_{i}}$};
  \draw (0.3,0.4) node [black]  {$\mathcal{F}$};
  \draw (1.06,0.25) node [black]  {$\mathcal{R}$};
  \node at (0.29,0.08) [circle,draw=gray,fill=gray!60]{};
  \node at (1.05,0.08) [circle,draw=gray,fill=gray!60]{};
  \draw[->,gray!60,very thick] (0.37,0.08)--(1.05,0.08);
  \draw[->,gray!60,very thick] (0.29,0.08)--(0.5,0.08);
  \draw[->,gray!60,very thick] (0.37,0.08)--(0.7,0.08);
  \draw[->,gray!60,very thick] (0.37,0.08)--(0.9,0.08);
  \draw[->,gray!60,very thick] (0.58,0.2)--(1,0.2);
  \draw[->,gray!60,very thick] (0.58,0.2)--(0.84,0.2);
  \draw[->,gray!60,very thick] (0.51,0.2)--(0.62,0.2);
  \draw[gray!60,very thick] (0.7,0.32)--(0.91,0.32);
  \draw[->,gray!60,very thick] (0.7,0.32)--(0.8,0.32);
\end{tikzpicture}
\caption{The energy $U$ of the system as a functional of the field $\phi$. At $t=0$, the system can be described as a thermal ensemble of states inside the basin $\mathcal{F}$, which is separated from $\mathcal{R}$ by a local maximum along the tunneling trajectory in field space. The field-space hypersurface $\Sigma_{\varphi_i}$ contains all field configurations in $\mathcal{R}$ that are energetically degenerate with $\varphi_i$. $\phi_{\rm top}$ is the familiar sphaleron.
}
\label{fig:FRBarrier}
\end{figure}

    

{\it \bf Evaluating the Tunneling Rate.} Assuming the field's initial state to be fully localized within $\mathcal F$, we describe the system through
\begin{align}\label{eq:IniCond}
	\langle \varphi_i | \rho | \varphi_j \rangle = 
 \begin{cases}
     \langle \varphi_i |e^{- \beta H} | \varphi_j \rangle \ {\rm if} \ \varphi_i, \varphi_j \in {\mathcal F}, \\ 
     0 \ {\rm otherwise}.
 \end{cases}
\end{align}
This choice makes manifest a crucial conceptual subtlety affecting any attempt to even define the tunneling rate in a finite-temperature system. In principle, such a system would have to be described by a density matrix of the form $\rho \propto e^{-\beta H}$ on the entire Fock space. Such a state, however, would be stationary, implying that no tunneling would occur. Instead, Eq.~\eqref{eq:IniCond} can be understood to describe an idealized \textit{supercooled} state.

A possible interpretation of such a state is as a remnant of a state that had thermalized in a modified version of the potential, in which the true vacuum basin had been removed, e.g., through an external force or finite-temperature corrections.\footnote{This parallels the definition of the false vacuum state laid out in Ref.~\cite{Andreassen:2016cvx}. These authors argue that the false vacuum cannot be thought of as an energy eigenstate, precisely due to its stationary nature and large support in the true vacuum basin. Instead, the false vacuum state should be understood as superposition of so-called \textit{resonance states}, which can be approximated by the eigenfunctions of the potential upon removal of the true vacuum basin.} After the formation of the true vacuum we expect an initial sloshing behavior, corresponding to high-energy excitations within the ensemble classically propagating outside of the basin, analogous to the zero-temperature case discussed in Ref.~\cite{Andreassen:2016cvx}. Given the strong dependence of these transient effects on the concrete system and complexity of the sloshing dynamics, we therefore restrict ourselves to an analysis of the tunneling rate \textit{after} the state has been prepared. The applicability of our analysis is therefore controlled by the accuracy with which the state of a system can be described by Eq.~\eqref{eq:IniCond} at the chosen initial time $t=0$.

To evaluate Eq.~\eqref{PR1}, we rewrite the density matrix as
\begin{align}\label{rho*}
    \langle\varphi_{i} |\rho|\varphi_j \rangle =& \sint{\mathcal{F}}{}{0} D \varphi_* \langle\varphi_{i} |e^{-\beta_1 H} |\varphi_* \rangle  \langle\varphi_* | e^{-\beta_2 H}  |\varphi_j \rangle  \\ 
    = \int_{\mathcal{F}} D \varphi_* &\sint{\varphi_1(0)=\varphi_i}{\varphi_1 (\beta_1)=\varphi_*}{0.9}  \mathcal{D} \varphi_1 \  e^{-S_E [\varphi_1]} \sint{\varphi_2(0)=\varphi_*}{\varphi_2(\beta_2)=\varphi_j}{0.9}  \mathcal{D} \varphi_2 \  e^{-S_E [\varphi_2]} \\ 
    \equiv  \int_{\mathcal{F}} D \varphi_* & E ( \varphi_i| \varphi_*,\beta_1 ) E (\varphi_*,\beta_2|  \varphi_j ),
\end{align}
with $\beta = \beta_1 + \beta_2$. In the last line, we introduced the Euclidean-time propagator $E$.

We also rewrite the matrix elements representing a real-time evolution in terms of propagators. Using that $| \varphi_{i,j} \rangle$ and $|\varphi_f\rangle$ lie in $\mathcal F$ and $\mathcal R$, respectively, we have~\cite{Andreassen:2016cff,Andreassen:2016cvx}
\begin{align}
    D_F & (\varphi_j | \varphi_f ,t ) \equiv  \langle \varphi_j | \varphi_f ,t_f \rangle = \nonumber \\
    &= \int_{\lowersub\Sigma_{\varphi_j}} \hspace{-1em} D \sigma \int_0^t \text{d}s \ \bar{D}_F (\varphi_j | \sigma ,s ) D_F ( \sigma ,s  | \varphi_f ,t ). \label{Ddecomp} 
\end{align}
To define $\bar{D}_F$ we introduce the functional $T_{\varphi_k}[\varphi]$, which maps any field $\varphi$ onto the time when it first reaches the field-space hypersurface $\Sigma_{\varphi_k} \subset {\cal R}$ containing all the configurations in ${\cal R}$ with the same energy as $\varphi_k$. In terms of this object, $\bar{D}_F$ is defined as
\begin{equation}
    \bar{D}_F (\varphi_i | \sigma ,s ) \equiv \sint{\varphi (0) = \varphi_i}{\varphi (s) = \sigma}{1}  \mathcal{D} \varphi \ e^{i S[\varphi]} \delta \left( T_{\varphi_i}[\varphi] -s \right).
\end{equation}
$T_{\varphi_i}[\varphi] = s$ is the \textit{crossing condition}. In the relevant limit of physical times large relative to the natural time scale for motions within the potential, we take $T_{\varphi_*}[\varphi] \simeq T_{\varphi_i}[\varphi]$ for all $\varphi_i \in \mathcal{F}$.

Note that our anlysis relies on no additional assumptions regarding the dynamics once the initial state in Eq.~\eqref{eq:IniCond} has been prepared. In fact, as pointed out in the beginning of this section, this would correspond to a stationary density matrix without tunneling. Instead, we only impose that the system be in the supercooled state described by Eq.~\eqref{eq:IniCond} at $t=0$, and allow for entirely free evolution of the system afterwards.

Eq.~\eqref{Ddecomp} amounts to splitting the time evolution from $\varphi_i$ to $\varphi_f$ into a piece connecting $\varphi_i$ with some $\sigma$ in $\Sigma_{\varphi_i}$, and the time evolution thereafter. Using the same representation for $D_F  ( \varphi_f ,t  |\varphi_j)$, we rewrite $P_{ji}$ as
\begin{align}
    P_{ji}=& \int_{\lowersub\Sigma_{\varphi_*}} 
    \hspace{-1em} D \sigma \int_0^t \hspace{-0.4em}\text{d}s   \int_{\lowersub\Sigma_{\varphi_*}} \hspace{-1em} D \sigma^{\prime} \int_0^t \hspace{-0.4em}\text{d}s^{\prime} \,  \bar{D}_F (\varphi_j | \sigma ,s ) \bar{D}_F^* (\varphi_i | \sigma^\prime ,s^\prime ) \nonumber \\
    & \times \int_{\lowersub \cal R} \hspace{-0.4em} D \varphi_f \, D_F(\sigma, s | \varphi_f,t)  D_F( \varphi_f,t  | \sigma^\prime, s^\prime)  .
    \label{Pij1}
\end{align}
Following Refs.~\cite{Andreassen:2016cff,Andreassen:2016cvx} we neglect back-tunneling, corresponding to $\int_{\cal R} D \varphi_f \vert \varphi_f \rangle \langle \varphi_f \vert \simeq 1$. 

The second line of Eq.~(\ref{Pij1}) thus evaluates to $D_F(\sigma, s | \sigma^\prime, s^\prime)$. Next we rewrite the time integrals as
\begin{equation}
    \int_0^t \text{d}s  \int_0^t \text{d}s^{\prime} = \int_0^t \text{d}s  \int_0^s \text{d}s^{\prime} + \int_0^t \text{d}s^{\prime} \int_0^{s^{\prime}} \text{d}s  .
\end{equation}
Now we recombine the propagator in Eq.~\eqref{Pij1} with one of the two factors of $\bar{D}_F$, using Eq.~\eqref{Ddecomp}. Eliminating the remaining time integral through a derivative, we find 
\begin{align}
    \frac{{\rm d} P_{ji}}{ {\rm d} t}= \int_{\lowersub \Sigma_{\varphi_*}} \hspace{-1em} D \sigma \ \bar{D}_F(\varphi_j|\sigma, t) D_F^* (\sigma, t | \varphi_i) + c.c.
\end{align}
As $\rho_{ij}$ is time-independent, $\Gamma$ thus becomes
\begin{widetext}
   \begin{align}
    \Gamma=& \frac{1}{\mathcal{P}_{\mathcal F}(t)} \sint{\mathcal{F}}{}{0}  D \varphi_*  \sint{\Sigma_{\varphi_*}}{}{0}  D \sigma  \sint{\mathcal{F}}{}{0}  D \varphi_i \sint{\mathcal{F}}{}{0}  D \varphi_j E ( \varphi_i| \varphi_*,\beta_1 ) E (\varphi_*,\beta_2|  \varphi_j )  \bar{D}_F(\varphi_j|\sigma, t) D_F(\sigma, t | \varphi_i) + c.c.  \label{GammaMaster}
\end{align}
\end{widetext}
Similar to the integral over the final state, we approximate the integrals over $\varphi_{i,j}$ as $\int_{\mathcal{F}} D \varphi_{i,j} |\varphi_{i,j} \rangle \langle \varphi_{i,j}| \simeq 1$. The restriction on $\varphi_{i}$ is preserved by extending the crossing condition to the Euclidean-time propagators. Integrating over $\varphi_{i,j}$ allows us to merge each of the real-time path integrals in Eq.~\eqref{GammaMaster} with its ``neighboring'' Euclidean-time integral. Each of the remaining integrals can be evaluated through a deformation of the complex-time integration contour, see Fig.~\ref{fig:SKcontour}.

\begin{figure}[h!]
\begin{tikzpicture}[scale=1.9]
    \filldraw [gray!60] (0,-0.35) circle [radius=1pt] 
                     (3.6,0) circle [radius=1pt] ;
    \filldraw [gray!60] (0,-0.05) circle [radius=1pt];
    \filldraw [gray!60] (0,0.05) circle [radius=1pt];
     \filldraw [gray!60] (0,0.35) circle [radius=1pt];
    \draw[->](0,0) -- (3.9,0);
    \draw[->](0,-0.5)--(0,0.7);
    \draw (4.2,0) node [black]
                       {Re($z$)};
    \draw (0,0.9) node [black] 
                       {Im($z$)};
    \draw (0.33,0.2) node [black]
                       {$\varphi_{j}$};
    \draw (0.18,0.5) node [black]
                       {$\varphi_{*}$};
    \draw (0.18,-0.5) node [black]
                       {$\varphi_{*}$};
    \draw (0.33,-0.2) node [black]
                       {$\varphi_{i}$};
    \draw (-0.15,-0.2) node [black]
                       {$\beta_1$};
    \draw (-0.15,0.2) node [black]
                       {$\beta_2$};
    \draw[gray!60](0,-0.05)--(0.23,-0.16);
    \draw[gray!60](0,0.05)--(0.23,0.16);
    \draw[gray!60](3.6,0)--(3.7,-0.16);
    \draw (4,-0.2) node [black]
                       {$\sigma\in\Sigma_{\varphi_{j}}$};
    \draw[->,red!60,ultra thick] (0,0.35)--(2.4,0.15);
    \draw[->,red!60,ultra thick] (0,0.35)--(1.2,0.25);
    \draw[->,red!60,ultra thick] (0,0.35)--(3.6,0.05);
    \draw[->,gray!60,ultra thick] (0,-0.05)--(0,-0.35); 
    \draw[->,gray!60,ultra thick] (0,0.35)--(0,0.05);
    \draw[->,red!60,ultra thick] (3.6,-0.05)--(2.4,-0.15);
    \draw[->,red!60,ultra thick] (3.6,-0.05)--(1.2,-0.25);
    \draw[->,red!60,ultra thick] (3.6,-0.05)--(0,-0.35);
    \draw[->,gray!60,ultra thick] (0,0.05)--(2.4,0.05);
    \draw[->,gray!60,ultra thick] (0,0.05)--(1.2,0.05);
    \draw[->,gray!60,ultra thick] (0,0.05)--(3.6,0.05);
    \draw[->,gray!60,ultra thick] (3.6,-0.05)--(2.4,-0.05);
    \draw[->,gray!60,ultra thick] (3.6,-0.05)--(1.2,-0.05);
    \draw[->,gray!60,ultra thick] (3.6,-0.05)--(0,-0.05);
    \draw[->,blue!60,ultra thick] (3.6,-0.05)--(3.6,-0.35); 
    \draw[->,blue!60,ultra thick] (3.6,0.35)--(3.6,0.05);
    \draw[->,blue!60,ultra thick] (0,0.35)--(2.4,0.35);
    \draw[->,blue!60,ultra thick] (0,0.35)--(1.2,0.35);
    \draw[->,blue!60,ultra thick] (0,0.35)--(3.6,0.35);
    \draw[->,blue!60,ultra thick] (3.6,-0.35)--(2.4,-0.35);
    \draw[->,blue!60,ultra thick] (3.6,-0.35)--(1.2,-0.35);
    \draw[->,blue!60,ultra thick] (3.6,-0.35)--(0,-0.35);
\end{tikzpicture}
\caption{Gray: The relevant contours for Eq.~\eqref{GammaMaster}. Red: After the integral over $\varphi_{i,j}$ has been performed, the contour can be deformed in the complex plane. Blue: The contour along which the complex action can be evaluated most easily.}
\label{fig:SKcontour}
\end{figure}

For large times $t$, the diagonal contours in Fig.~\ref{fig:SKcontour} amount to infinitesimal Wick-rotation by the angles $\epsilon_{1,2}=\beta_{1,2}/t$. The dynamics along this contour can be described by a propagator of the usual form, albeit with an action defined with respect to the new, complex time variable $z = (1 - i \epsilon) t$. This yields  
\begin{align}
    \Gamma=& \frac{1}{\mathcal{P}_{\mathcal F}(t)} \sint{\mathcal{F}}{}{0}  D \varphi_*  \hspace{-0.4em}\sint{\Sigma_{\varphi_*}}{}{0}  D \sigma   \bar{D}_{F}^{\epsilon_2}(\varphi_*|\sigma, t) D_{F}^{\epsilon_1}(\sigma, t | \varphi_* ) + c.c. \nonumber \\ 
    =& \frac{1}{\mathcal{P}_{\mathcal F}(t)} \int_{\mathcal F} D \varphi_*  \sint{\Sigma_{\varphi_*}}{\ }{0}  D \sigma  \sint{\varphi_1 (0)=\varphi_*}{\varphi_1 (t)=\sigma}{1.2} \mathcal{D} \varphi_1 \ e^{i S_{\epsilon_1}[\varphi_1] } \delta \left( T_{\sigma}[\varphi_1] - t \right)  \nonumber \\ 
    &\times \left(\hspace{1.2 em}\sint{\varphi_2 (0)=\varphi_*}{\varphi_2 (t)=\sigma}{1.2} \mathcal{D} \varphi_2 \  e^{ iS_{\epsilon_2}[\varphi_2]} \right)^* + c.c.\label{GammaMasterDiag}
\end{align}
All quantities with a label $\epsilon_{1,2}$ are defined with respect to the complex time variable $z_{1,2}=(1-i \epsilon_{1,2})t$. 

\begin{figure}[h!]
    \centering
    \includegraphics[width=3.1in]{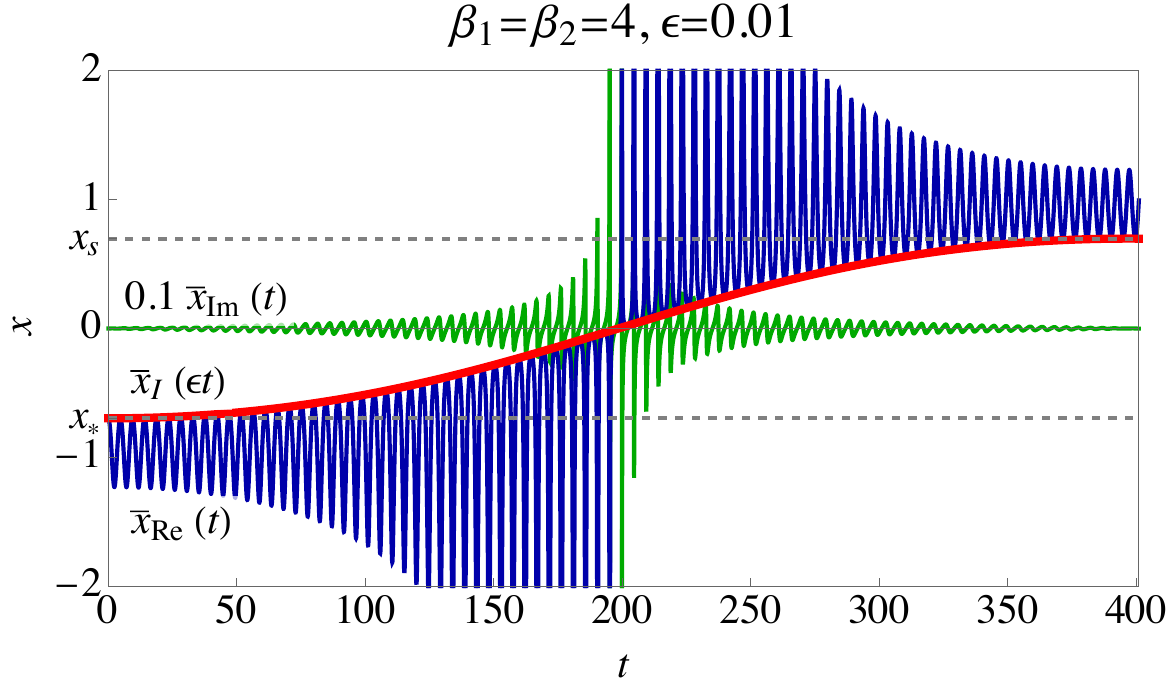}
    \caption{A steadyon solution for a point particle in a double well potential, $V(x)= \frac{1}{4}(x-1)^2$. Blue: The real part of the position. Green: The imaginary part. Red: The turning points of the real part constitute the Euclidean-time instanton.}
    \label{fig:steadyonl}
\end{figure}

\textbf{Steadyons.} Eq.~(\ref{GammaMasterDiag}) can be evaluated through a stationary phase approximation. We refer to the corresponding saddle points as $\phi_{1,2}$, and their actions as $S_{1,2}$. While the boundary conditions of all path integrals imply real values of the field at the initial and final times, the complex factors in the equation of motion (\textit{e.o.m.}) lead to complex solutions for intermediate times, so-called \textit{steadyons}~\cite{Steingasser:2024ikl}. By splitting the field into its real and imaginary parts, the steadyons describe the dynamics of two real fields linked through an interaction controlled by the infinitesimal parameter $\epsilon_{1,2}$. For small times, this implies that the real part of the field essentially undergoes its classical motion within $\mathcal{F}$, steadily sourcing the imaginary part, which in turn drives the motion of the real part, thus allowing it to reach $\mathcal{R}$. See Fig.~\ref{fig:steadyonl} for an example. The e.o.m.~suggests that $\phi_{1,2}$ are analytical functions on their respective slices of the complex-time plane. Furthermore, the arbitrariness of the decomposition $\beta= \beta_1 + \beta_2$ and the integral over $\varphi_*$ suggest that both of these solutions can be combined into a single function with $\beta$-periodicity in the Euclidean-time direction. See Fig.~\ref{fig:EuclideanInstanton}.

The normalization factor $\mathcal{P}_{\mathcal F}(t)$ is dominated by the constant saddle point $\phi_{\rm FV}$ (see Fig.~\ref{fig:FRBarrier}), leading to a vanishing exponent. $\Gamma$ is thus to leading order of the form
\begin{align}
    \Gamma \simeq & A \cdot e^{i S_1 - iS^*_2}  + c.c. \nonumber \\ 
    =& A \cdot e^{- \left( {\rm Im}  S_1 + {\rm Im}  S_2  \right)  + i \left( {\rm Re}  S_1  - {\rm Re}  S_2  \right)  }+ c.c.
    \label{GammaMasterSP}
\end{align}
The steadyons' e.o.m.~imply that their projection onto contours parallel to the real- and Euclidean-time axis are solutions of the e.o.m.~with respect to their corresponding time variables. First, this implies that the projections of the steadyons onto real-time contours starting from $\varphi_*$ describes the classical, real-time evolution of this configuration. Thus, the contribution of these pieces of the contours to the overall action would be strictly real, and identical for the upper and lower contour, such that the complex phase in Eq.~\eqref{GammaMasterSP} cancels. Similarly, the projections of the steadyons onto Euclidean-time contours that end on such configurations describe the ``classical'' motion of the configuration in imaginary time. Together with the $\beta$-periodicity, this implies that projections of the steadyons onto the Euclidean-time axis, or the contour parallel to it below $\sigma$, are solutions of the Euclidean equations of motion. These two solutions can be identified with pieces of a single, $\beta$-periodic instanton, see Fig.~\ref{fig:EuclideanInstanton}. Together with the factor $i$ from the integration contour, the total exponent is thus given by the Euclidean action of the $\beta$-periodic instanton.

\begin{figure}
\begin{tikzpicture}[scale=1.5]
  \draw[->] (-1.3,-0.5) -- (1.45,-0.5) node[right] {$x$};
  \draw[->] (0,-0.5) -- (0,0.8) node[above] {$V(x$)};
  \draw[domain=-1.4:1.4,smooth,variable=\x,very thick] plot ({\x},{cos(3*\x r)/2});  
  \filldraw [gray!60] (-0.52,-0.1) circle [radius=1pt]
                      (0.52,-0.1) circle [radius=1pt] ;
  \draw[->, gray!60,very thick](-0.52,-0.1)--(-0.33,0.2);
  \draw[->, gray!60,very thick](0.52,-0.1)--(0.33,0.2);
  \draw[->, gray!60,very thick](-0.33,0.2)--(-0.1,0.43);
  \draw[->, gray!60,very thick](0.33,0.2)--(0.1,0.43);
  \draw[thick,dashed] (-0.3,-0.5)-- (-0.3,0.27);
  \draw[thick,dashed] (0.3,-0.5)-- (0.3,0.27);
  \draw[thick,dashed] (-0.3,0.27)-- (0.3,0.27);
  \draw (-0.3,-0.7) node [black]  {$x_{*}$};
  \draw (0.3,-0.7) node [black]  {$\sigma$};
  \draw [gray!60,very thick](-0.1,0.43) arc [start angle=140, end angle=40, radius=0.13];
  \draw[->] (2,0) -- (3.55,0) node[right] {$\tau$};
  \draw[->] (2,-1.1) -- (2,1.1) node[above] {$x$};
  \draw[domain=2:3.55,smooth,variable=\x,very thick] plot ({\x},{-sin(5*(\x-2) r)});  
  \draw (1.9,0.65) node [black]  {$\sigma$};
   \draw (2.45,0.2) node [black]  {$\beta_{1}$};
  \draw (3.05,0.2) node [black]  {$\beta_{2}$};
  \draw[thick,dashed] (2.15,-0.65)-- (2.15,0);
  \draw[thick,dashed] (2.77,0.65)-- (2.77,0);
  \draw[thick,dashed] (2,-0.65)-- (3.4,-0.65);
   \draw[thick,dashed] (2,0.65)-- (2.75,0.65);
   \draw[blue, thick](2.15,-0.07)--(2.15,0.07);
  \draw[blue, thick](2.77,-0.07)--(2.77,0.07);
  \draw[blue, thick](3.4,-0.07)--(3.4,0.07);
  \draw (1.9,-0.65) node [black]  {$x_{*}$};
\end{tikzpicture}
\caption{The periodic Euclidean-time instanton that determines the tunneling rate in the limit $t \to \infty$ for a point particle in a symmetric double-well potential $V(x)$. The light-gray arrows indicate the periodic motion of the system within the inverted potential $ - V(x)$. The saddle-point values of $x_*$ and $\sigma$, necessary for the existence of a solution, depend on the partition of $\beta$ into $\beta_1$ and $\beta_2$.}
\label{fig:EuclideanInstanton}
\end{figure}

The corresponding steadyon does \textit{not} describe a physical process, as the precise chronology of such a process would depend on the chosen complex-time contour. It is rather a formal tool to evaluate the sum in Eq.~\eqref{GammaMasterSP}, which inherits the structure of the individual contributions. A possible interpretation of $\varphi_*$ as the dominant departure point of the tunneling is inconsistent with the boundary conditions; in Ref.~\cite{Steingasser:2024ikl}, it is argued that the tunneling out of an eigenstate is typically dominated by an instanton with initial momentum rather than a periodic configuration. Similarly, in the finite-temperature case, the $\beta$-periodicity arises from summing over the thermal ensemble. In Ref.~\cite{Steingasser:2024ikl} it is also pointed out that the description of the real-time contributions to Eq.~\eqref{GammaMasterSP} in terms of steadyons requires a regularization of the theory, amounting to infinitesimal Wick rotations of the real-time contours. See Fig.~\ref{fig:SKcontourIndividual}. Thus, the actual rate is determined by processes corresponding to paths extending over a variety of Euclidean-time intervals, each larger than $\beta$. The necessity of this regularization also affects the finite-temperature steadyon, which satisfies the crossing condition along any complex-time contour, except those overlapping with the real-time axis. These paths are a special case of a Lefschetz thimble, indicating that our analysis can be considered an application of Picard-Lefschetz theory~\cite{Witten:2010cx,Tanizaki:2014xba,Cherman:2014sba,Dunne:2015eaa,Bramberger:2016yog,Michel:2019nwa,Mou:2019gyl,Hertzberg:2019wgx,Ai:2019fri,Hayashi:2021kro,Nishimura:2023dky}. 

\begin{figure}[h!]
\begin{tikzpicture}[scale=1.7]
    \filldraw [gray!60] (0,-0.83) circle [radius=1pt] 
                     (3.6,0) circle [radius=1pt] ;
    \filldraw [gray!60] (0,-0.3) circle [radius=1pt];
    \filldraw [gray!60] (0,0.55) circle [radius=1pt];
     \filldraw [gray!60] (0,0.285) circle [radius=1pt];
    \draw[->](0,0) -- (3.9,0);
    \draw[->](0,-0.8)--(0,0.7);
    \draw (4.2,0) node [black]
                       {Re($z$)};
    \draw (0,0.9) node [black] 
                       {Im($z$)};
    \draw (0.3,0.45) node [black]
                       {$\varphi_{j}$};
    \draw (0.18,0.65) node [black]
                       {$\varphi_{*}$};
    \draw (0.18,-0.85) node [black]
                       {$\varphi_{*}$};
    \draw (0.3,-0.45) node [black]
                       {$\varphi_{i}$};
    \draw (-0.15,-0.6) node [black]
                       {$\beta_1$};
    \draw (-0.15,0.45) node [black]
                       {$\beta_2$};
    \draw (-0.15,-0.15) node [black]
                       {$\tau_i$};
    \draw (-0.15,0.1) node [black]
                       {$\tau_j$};
    \draw[gray!60](0,-0.32)--(0.23,-0.43);
    \draw[gray!60](0,0.32)--(0.23,0.43);
    \draw (3.65,-0.2) node [black]
                       {$\sigma\in\Sigma_{\varphi_{j}}$};
    \draw[->,gray!60,ultra thick] (0,0.3)--(2.4,0.1);
    \draw[->,gray!60,ultra thick] (0,0.3)--(1.2,0.2);
    \draw[->,gray!60,ultra thick] (0,0.3)--(3.6,0);
    \draw[->,gray!60,ultra thick] (0,-0.38)--(0,-0.58);
    \draw[->,gray!60,ultra thick] (0,-0.38)--(0,-0.78); 
    \draw[->,gray!60,ultra thick] (0,0.55)--(0,0.31);
    \draw[->,gray!60,ultra thick] (3.6,0)--(2.4,-0.1);
    \draw[->,gray!60,ultra thick] (3.6,0)--(1.2,-0.2);
    \draw[->,gray!60,ultra thick] (3.6,0)--(0,-0.3);
\end{tikzpicture}
\caption{The Keldysh-Schwinger contour representing one of the contributions in Eq.~\eqref{GammaMasterSP}. The existence of a steadyon along the real-time branch requires a regularization which can be understood as an infinitesimal Wick rotation, extending over the imaginary-time interval.}
\label{fig:SKcontourIndividual}
\end{figure}

{\it \bf Multifield Models and Effective Action.}
In many applications the effects of external degrees of freedom affecting the tunneling through interactions are crucial~\cite{Moss:1985ve,Bodeker:1993kj,Quiros:1994dr,Salvio:2016mvj}. We can incorporate them into our discussion by extending $\mathcal{P}_\mathcal{R}$ as
\begin{equation}\label{fundgen}
    \mathcal{P}_{\mathcal R} (t)= \mathcal{N} \int D \chi \int_{\mathcal R} D \varphi \,\langle \varphi , \chi ,t | \rho | \varphi , \chi  ,t \rangle .
\end{equation}
For the zero-temperature case, it was argued in Ref.~\cite{Andreassen:2016cvx} that the proper way to account for these additional path integrals is to first solve for the instanton using the tree-level potential and only afterwards integrate over the remaining fields, such that their effect is reduced to additional contributions to the prefactor $A$ in Eq.~\eqref{GammaMasterSP}.

Loop corrections to the potential can be significant in finite-temperature systems, e.g., through particles obtaining thermal masses~\cite{Carrington:1991hz,Kajantie:1995dw,Salvio:2016mvj}. This suggests to integrate out the external fields \textit{before} performing the saddle-point approximation in order to capture these effects, which would amount to working with an effective action. In Ref.~\cite{Andreassen:2016cvx}, however, it was pointed out that the instanton background can spoil the convergence of momentum-dependent corrections to the effective action. Denoting the field giving rise to the instanton by $\phi$, these are suppressed by increasingly higher orders of $ m_{\phi}^2/m_i^2$, where $m_i^2$ is the field-dependent mass of the particles being integrated out and evaluated on the instanton~\cite{Salvio:2016mvj,Moss:1985ve,Bodeker:1993kj}. Hence, while the fluctuations of the field $\phi$ always need to be evaluated in their functional determinant form, any other field can be accounted for through its contributions to the effective action as long as its (effective) mass is larger than that of the scalar field around the instanton~\cite{Gleiser:1993hf}. When this ratio is spacetime-dependent, the expansion is controlled by its value in the spatial region where the instanton is localized, as the Euclidean action predominantly depends on its values in this region. If the tunneling is only made possible by loop contributions from a particle which cannot be integrated out, such as the scalar field itself, neither this approach nor the strategy suggested in Ref.~\cite{Andreassen:2016cvx} is applicable.

\begin{figure}[t]
    \centering
    \includegraphics[width=3.1in]{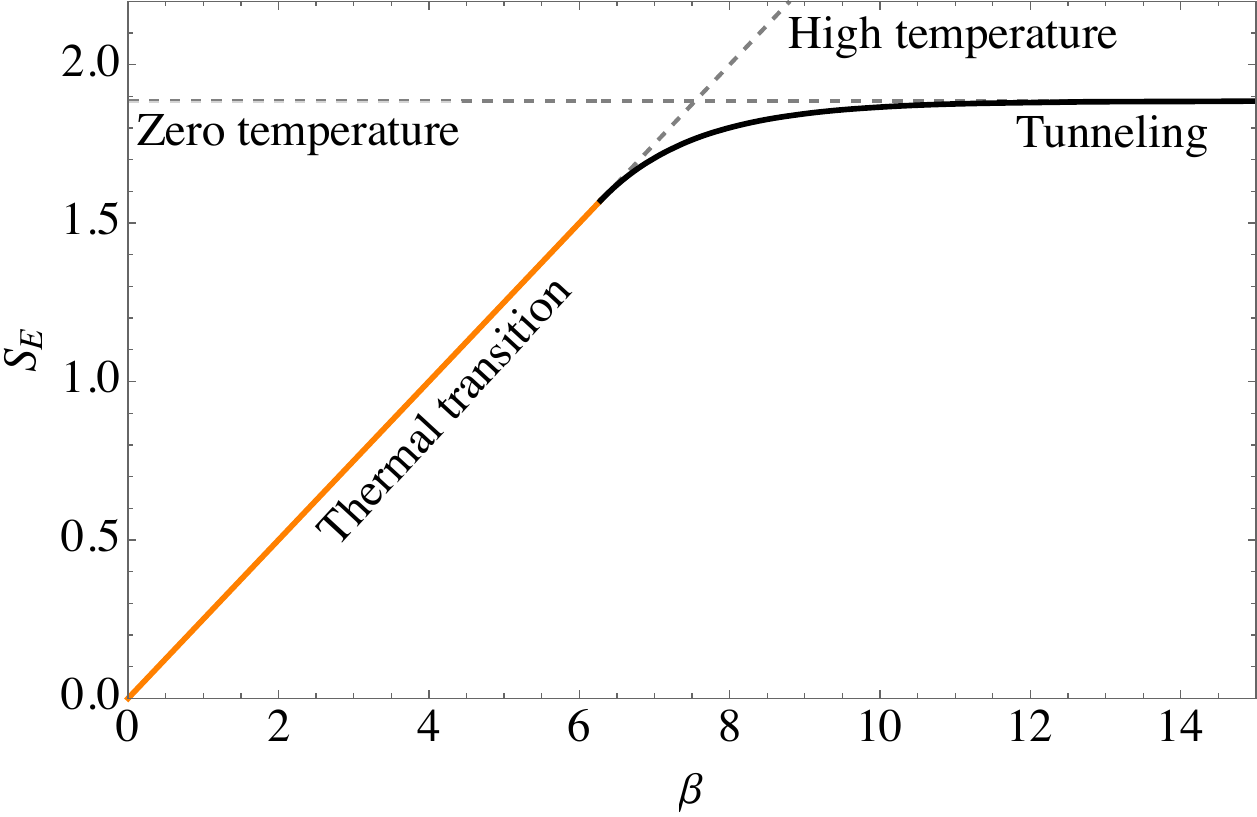}
    \caption{The Euclidean action as a function of inverse temperature $\beta$ for the potential $V(x)= \frac{1}{4}(x^2-1)^2$. Quantum tunneling requires $\beta \gtrsim 6.28$ in these units, which corresponds to the oscillator period around the minimum of the inverted potential. For smaller values of $\beta$, the transition is entirely determined by thermal excitations.}
\label{figActionPlot}
\end{figure}

{\it \bf Limits.}
For low temperatures, $\beta \to \infty$, we recover the familiar zero-temperature results. As the high-temperature limit corresponds to large occupation numbers, it can be expected to reproduce the classical transition rate~\cite{Langer:1967ax,Langer:1969bc,Bochkarev:1992rh,Boyanovsky:1993ki,Garny:2012cg,Ekstedt:2022tqk,Hirvonen:2021zej,Lofgren:2021ogg}, where the exponent is to leading order determined by a constant solution in the Euclidean-time direction, the \textit{sphaleron}. See Fig.~\ref{figActionPlot}.

For a point particle, the finite-temperature instanton represents a periodic motion in the inverse potential. For generic potentials, the Euclidean time necessary for such motions is bounded from below by some $\Delta \tau_{\rm min}$ of order of the potential's typical time scale. For smaller values of $\beta$, no periodic solution exists, leaving only the solution corresponding to the particle being at rest on top of the potential barrier. This configuration on its own violates the crossing condition, but infinitesimal deformations of it still serve as approximate solutions. 

Non-perturbative effects can influence this limit. Integrating out heavier fields and resumming so-called ring diagrams can induce temperature-dependent corrections to the potential, including a mass term. For sufficiently large temperatures, the magnitude of these terms is controlled by the temperature alone. As an example, the thermal mass is generally of the form $m_T^2 = \kappa \cdot T^2$, with some combination $\kappa$ of the scalar field's couplings to the background fields, but, importantly, no additional loop-suppression factor $(4 \pi)^{-2 N_{\rm loop}}$. In other words, all relevant energy scales of the theory are of order of the temperature up to numerical coefficients, which can be ${\cal O}(1)$ for sufficiently large couplings. This can be understood as the energy per particle increasing due to its coupling to an increasingly hot background plasma counteracting the increase in the occupation number. An important example for this behavior is the Standard Model Higgs field, for which we find that the coefficient $\kappa$ lies within the range $0.1-0.2$ for all energies above the central value of the instability scale, $\mu_I\sim10^{11}$~GeV~\cite{Steingasser:2023ugv}. While this establishes in principle the possibility of a perturbative expansion, it also suggests that precision calculations should take into account leading-order corrections in $\kappa$, in particular since the effects of, e.g., potentially large right-handed neutrinos can further enhance this effect~\cite{Chauhan:2023pur}.

{\it \bf Discussion.}
We have derived a compact path-integral representation for the tunneling rate in a finite-temperature system. Its exponent can be determined through a stationary-phase approximation, which is dominated by a $\beta$-periodic steadyon. In the limit of large physical times, the exponent reduces to the Euclidean action of the instanton obtained by projecting this steadyon onto a Euclidean-time contour of length $\beta$.  

This simple relation explains how the tunneling rate, which is \textit{a priori} a real-time-dependent quantity, can be described in terms of a Euclidean-time quantity. Our analysis shows that the periodic instanton does \textit{not} represent the contribution of one dominant state within the thermal ensemble, but rather emerges as a formal tool for summing over the ensemble. In addition, we have analyzed the influence of background fields on the tunneling rate as well as subtleties related to both the high- and low-temperature limits. This establishes a robust foundation for tunneling and bubble-nucleation calculations at arbitrary temperatures.

\vskip 10pt

{\it Acknowledgements.} It is a pleasure to thank Alan H.~Guth for helpful discussions. TS's contributions to this work were made possible by the Walter Benjamin Programme of the Deutsche Forschungsgemeinschaft (DFG, German Research Foundation) -- 512630918. MK is supported in part by the MLK Visiting Scholars Program at MIT. Portions of this work were conducted in MIT's Center for Theoretical Physics and partially supported by the U.S. Department of Energy under Contract No.~DE-SC0012567. This project was also supported in part by the Black Hole Initiative at Harvard University, with support from the Gordon and Betty Moore Foundation and the John Templeton Foundation. The opinions expressed in this publication are those of the author(s) and do not necessarily reflect the views of these Foundations.

\bibliography{ref}

\end{document}